\documentstyle[twocolumn,prb,aps,epsf]{revtex}

\begin{document}

\draft

\title{Many-Body Vortex Effect on the Transverse Force
Acting on a Moving Vortex}

\author{Nobuhiko Hayashi}
\address{Computer Center, Okayama University, Okayama 700-8530, Japan}
\date{\today}

\maketitle

\begin{abstract}
   The nondissipative transverse force acting on
one moving vortex under the influence of another vortex is discussed
in fermionic superfluid systems,
where the relative velocity between the vortices is finite.
   On the basis of detailed numerical solutions of
the Bogoliubov-de Gennes equation, the Berry phase for an adiabatic motion
of the vortex line is examined for a two-vortex system.
   It is found that the detailed electronic structure of a vortex core
can affect the transverse force, without abandoning the previous
discussions of the robust Magnus force on a single vortex.
\end{abstract}


   Much interest has been devoted to the dynamical motion and pinning
effect of vortices
existing in various situations in nature
such as
type-II superconductors~\cite{super},
superfluidity in $^3$He, and
superfluid neutron star matters.~\cite{fabio,mochizuki}
   It is essential also from the applied physics point of view
that we understand transport
properties due to vortices in type-II superconductors,
especially for high-$T_{\rm c}$ cuprates,
to better control superconducting characteristics
under magnetic fields.
   In addition, the recent progress of experiments on $^3$He in a rotating
cylinder
shed new light on the fundamental
physics field where exciting analogies between the early Universe
and the superfluid $^3$He were examined.~\cite{bevan-n,eltsov}
   For these systems,
the dynamics of moving vortices is the focal point.

   In this letter, we consider a {\it many-body effect} between vortices
on the dynamics of a vortex.
   We will propose that a novel force should act on one moving vortex
due to surrounding another vortex when the relative
velocity between the vortices is finite.

   Currently, in the case of fermionic superfluids and superconductors,
the fundamental equation of motion for a moving vortex
is highly controversial.
   In particular, the term of the nondissipative transverse force
in the equation of motion,
i.e., the vortex velocity dependent part of the Magnus force
(per unit length of the vortex axis $z$),
$\alpha{\hat z} \times {\bf v}_L$,
is under intensive
debate.~\cite{ao-prlcom,kopnin-prlrep,hall-prlcom,wexler-prlrep1}
Here, ${\bf v}_L$ is the velocity of a vortex.

   Broadly speaking, in the single vortex case,
at present
there are two conflicting points of view on
the theory of the transverse force.
   One theory~\cite{kopnin76,volovik,feigelman,spect}
considers a momentum transfer from the (vortex) system under
consideration to a heat bath system
by a relaxation of the quasiparticles of
the vortex bound states~\cite{caroli} inside a vortex core.
   In this theory, called the spectral flow theory,~\cite{spect}
the coefficient of
the transverse force, $\alpha$, essentially depends on
the vortex bound states
(i.e., the electronic structure inside a vortex core)
in combination with the relaxation time $\tau$ of the quasiparticles.
   Experimental results on $^3$He in a rotating cylinder support
the spectral flow theory,~\cite{volovik96}
which is the crucial point of the $^3$He simulation of
the early Universe.~\cite{bevan-n}
   In contrast, the other theory~\cite{ao,thouless}
claims that the transverse force on
a moving vortex is a robust quantity which never depends on
such quasiparticle relaxation inside a vortex core.
   This ``robust" Magnus force does not depend on the details of
the vortex bound states inside a vortex core,
but only on the superfluid density
far from the core~\cite{thouless}
as long as a single vortex is considered.
   There also exist experimental results supporting
the robust Magnus force theory,~\cite{zhu}
and a negative experimental result
for the spectral flow theory is obtained there.
   Thus, the transverse force issue is controversial now,
and careful interpretation of the experimental results
by generalizing the single-vortex theories mentioned above
to more complicated situations
will be required to resolve it,
as noted in ref.~\onlinecite{wexler-prlrep1}.
   To advance our knowledge of vortex dynamics,
it is important to consider
{\it many-body effects} between vortices whose relative velocity is nonzero,
due to, for example, vortex pinning.

   The transverse force
is also a key point in the understanding of the Hall effect
in the mixed state in type-II superconductors,~\cite{dorsey}
especially for
the mysterious sign reversal of the Hall resistivity (conductivity)
observed experimentally
in several type-II superconductors.~\cite{hagen}
   A typical objection to the robust Magnus force theory has arisen from
small Hall angles observed in most transport experiments, because
the robust (or full large) Magnus force leads to
an extremely large Hall angle and to no sign reversal,
as long as the many-body effect is not considered.
   However, in real superconductors, the vortex pinning cannot be
avoided, and then it is expected that the relative velocity of
vortices is generally nonzero, leading to many-body effects
between vortices.
   Ao~\cite{ao95} previously proposed that the movement of
vortex vacancies in the pinned vortex lattice
as a many-body vortex effect
should lead to
a relevant result and the sign reversal,
while he maintained
the robust Magnus force acting on individual moving vortices.
   Recent transport experiments~\cite{kang-danna}
certainly revealed that the vortex pinning played
an important role in the Hall effect in some way.
   Therefore,
it is required to further investigate what type of {\it many-body effect}
can exist in various possible vortex situations, (e.g.,
in the plastic flow state, in which portions of the vortex lattice
move while other portions remain pinned~\cite{plastic}), i.e.,
situations in which the relative velocity of
vortices is generally nonzero.

   In addition, in high-$T_{\rm c}$ cuprate superconductors,
it was universally observed that the sign-reversal Hall effect existed
only in underdoped samples, but not in overdoped ones.~\cite{nagaoka}
   A hypothetical concept arises from this result;
the change of the electronic state due to the doping
could be related to the internal electronic structure
inside vortex cores
so that it affects the dynamic property of vortices.
Such an idea opens a new possibility of
relating the doping-dependent Hall effect~\cite{nagaoka}
to the electronic structure of
various vortex cores currently the focus of attention, e.g.,
cores with the antiferromagnetic moment,~\cite{aferro}
in $d_{x^2-y^2}+id_{xy}$ superconductivity,~\cite{franz}
and with doping dependence.~\cite{himeda}

   Motivated by the above concepts,
we investigate a contribution from
the electronic structure around a vortex core
to the transverse force acting on a moving vortex
{\it under the influence of another vortex} (e.g., pinned one).
   We consider such a {\it many-body effect} between vortices
whose relative velocity is nonzero, by discussing a two-vortex system.
   A Berry phase picked up by the system
for an adiabatic vortex motion is examined
to derive the coefficient of the transverse force
$\alpha{\hat z} \times {\bf v}_L$,
according to the Berry phase approach
to the vortex dynamics,~\cite{ao,gaitan,dziarmaga}
on the basis of
numerical wave functions around each vortex.
   It should be noted that the relaxation time $\tau$ is not included
in the present analysis, and therefore
the contribution of the vortex core which we propose
is independent of the spectral flow theory~\cite{spect}
which essentially depends on
$\tau$.
   We will consider a neutral system and rectilinear vortices
(or vortices in a two-dimensional system) with vorticity antiparallel
to ${\hat z}$.

   We will base our analysis on
the Bogoliubov-de Gennes (BdG) theory,~\cite{BdG}
which is the spatially inhomogeneous version of
the BCS theory.
   The system is described
in terms of the Bogoliubov wave function
$\bigl( u_j({\bf r})$, $v_j({\bf r}) \bigr)$.
   We start with the BdG equation
given, in a dimensionless form, by
\begin{eqnarray}
\pmatrix{
K & \Delta({\bf r}) \cr
\Delta^{\ast}({\bf r}) & -K \cr
}
{\hat \chi}_j({\bf r})
=
E_j {\hat \chi}_j({\bf r}), \quad
{\hat \chi}_{j}({\bf r}) =
\pmatrix{u_j({\bf r}) \cr v_j({\bf r}) \cr},
\label{eq:bdg}
\end{eqnarray}
where $K=-\nabla^{2}/2 k_{\rm F}\xi_0-\mu$,
$\mu$ is the chemical potential, and
$\xi_0$(=$v_{\rm F} / \Delta_0$) is the coherence length
[$\Delta_0$ is the uniform gap at zero temperature $T=0$,
$k_{\rm F}$ ($v_{\rm F}$) is the Fermi
wave number (velocity), and $\hbar=1$].
   In eq.\ (\ref{eq:bdg}),
the length (energy) is implicitly measured by $\xi_0$ ($\Delta_0$).
   The system is characterized by a parameter
$k_{\rm F}\xi_0$.~\cite{haya}
   The pair potential is self-consistently determined with
$\Delta({\bf r})=
g\sum_{|E_j|\leq \omega_{\rm D}}
u_j({\bf r})v^{\ast}_j({\bf r})\{1-2f(E_j)\}$.
Here, $g$ is the coupling constant and $\omega_{\rm D}$ the energy cutoff,
which are related by the BCS relation
via the transition temperature $T_{\rm c}$ and
the gap $\Delta_0$.
   We set $\omega_{\rm D}=20\Delta_0$.

   To obtain exact solutions ${\hat \chi}_j({\bf r})$
(i.e., electronic structure) around a vortex,
we numerically calculate eq.\ (\ref{eq:bdg})
under the following conditions for clarity,
as in ref.~\onlinecite{haya}.
(a) The system is a cylinder with radius $R$.
(b) The Fermi surface is cylindrical.
(c) The pairing has isotropic $s$-wave symmetry.
   Thus, the system has cylindrical symmetry.
   We write the eigenfunctions as
$u_j({\bf r})=u_{n,l}(r) \exp\bigr[i(l-\frac{1}{2})\theta \bigl]$ and
$v_j({\bf r})=v_{n,l}(r) \exp\bigr[i(l+\frac{1}{2})\theta \bigl]$ with
$\Delta({\bf r})=\Delta(r) \exp\bigr[-i\theta \bigl]$
in cylindrical coordinates, where $n$ is the radial quantum number and
the angular momentum
$|l|=\frac{1}{2},\frac{3}{2},\frac{5}{2},\cdots$.
   We expand the eigenfunctions in terms of
the Bessel functions ~\cite{caroli}
$J_m(r)$ as ~\cite{gygi}
$u_{n,l}(r)=\sum_{i} c_{ni}\phi_{i|l-\frac{1}{2}|}(r)$,
$v_{n,l}(r)=\sum_{i} d_{ni}\phi_{i|l+\frac{1}{2}|}(r)$,
where $\phi_{im}(r)=[{\sqrt 2}/ RJ_{m+1}(\alpha_{im})]
J_m(\alpha_{im}r/ R)$,
$\alpha_{im}$ is the $i$-th
zero of $J_m(r)$, and $i=1,2,\cdots, N$.
   We set $R=20$--40 $\xi_0$.
   Equation (\ref{eq:bdg}) is reduced to a $2N\times 2N$
matrix eigenvalue problem.
   This useful technique to solve eq.\ (\ref{eq:bdg}),
developed by Gygi and Schl\"uter,~\cite{gygi}
has been utilized in many cases.~\cite{fabio,franz,haya,haya-j}

   Adiabatic vortex motion
produces a Berry phase $\phi_j$ in the solution
of eq.\ (\ref{eq:bdg}),
${\hat \chi}_j \rightarrow \exp[i\phi_j]{\hat \chi}_j$,
and the total Berry phase $\Gamma$ picked up by the whole system,
$|\Phi \rangle \rightarrow \exp[i\Gamma]|\Phi \rangle$,
is given as
$\Gamma=
-\sum_{j}
\phi_j$~\cite{gaitan,dziarmaga}.

   We consider the single vortex case first.
   When the center of the vortex ${\bf r}_{0}(t)$ adiabatically moves,
the motion gives rise to the Berry phase $\phi_j$ picked up by
the Bogoliubov wave function ${\hat \chi}_j$.
   With the cylindrical symmetry of
the vortex,
the total Berry phase is obtained as~\cite{gaitan,dziarmaga}
\begin{eqnarray}
\Gamma&=&
-\int dt \int d^{2}x
({\dot{\bf r}}_0 {\bf \nabla}_{{\bf r}_0} \theta) S(r) \nonumber \\
&=&-\pi \bigl[ S(\infty)-S(0) \bigr]
\int dt (x_0 {\dot y}_0 - y_0 {\dot x}_0),
\label{eq:single-berry}
\end{eqnarray}
where $S(r)$ is the canonical angular momentum
density composed of ${\hat \chi}_j$,
and ${\dot{\bf r}}_0={\bf v}_L$.
   In the framework of the BdG theory,
the physical quantity $S(r)$ must be expressed
{\it in the form with the Fermi distribution function} $f(E)$,
as pointed out by Gaitan,~\cite{gaitan} namely
\begin{eqnarray}
S(r) &=&
2 \sum_{E_{n,l} > 0}\Bigl[
-f(E_{n,l})|u_{n,l}(r)|^2 \Bigl( -l+\frac{1}{2} \Bigr) \nonumber \\
     & & {}+\{1-f(E_{n,l})\}|v_{n,l}(r)|^2 \Bigl( -l-\frac{1}{2} \Bigr)
\Bigr].
\label{eq:ang-s}
\end{eqnarray}

\begin{figure}
\epsfxsize=77mm
\hspace{5.5mm}
\epsfbox{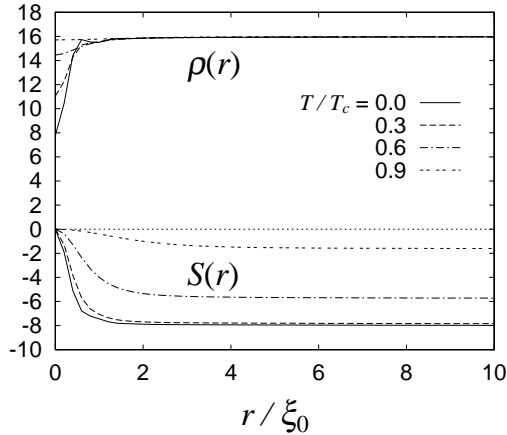}
\vspace{5.5mm}
\caption{
        The canonical angular momentum density $S(r)$
     and the total particle density $\rho(r)$ around a vortex
     at several temperatures,
     in arbitrary units, but on a common scale.
        $k_{\rm F}\xi_0=4$.
        Note the relation $S=-\rho/2$ at $T=0$ far from
     the vortex center $r=0$.
}
\label{fig:1}
\end{figure}

   At $T=0$, eq.\ (\ref{eq:ang-s}) which is
calculated using our numerical solutions $(u_j,v_j)$
of eq.\ (\ref{eq:bdg}) gives $S(0)=0$ and
the uniform value $S=-\rho/2$ far from the vortex core
(see Fig.\ \ref{fig:1}),
where $\rho/2$ is half the total particle density
and is, at $T=0$, equal to
the density of the Cooper pair,
i.e., half the superfluid density $\rho_s/2$.
   Thus, the expression for the Magnus force derived
at $T=0$ by Ao and Thouless,~\cite{ao}
$\alpha=-h\rho_s/2$,
is obtained from eqs.\ (\ref{eq:single-berry})
and (\ref{eq:ang-s})
as in refs.~\onlinecite{gaitan} and \onlinecite{dziarmaga}.
   In the present study, we extend the above
discussion for $T=0$ to
the finite temperature case
within the mean-field BdG framework and
with an ansatz that, at finite temperatures,
the total Berry phase is the sum
of the thermally weighted contribution from
each Bogoliubov wave function, i.e.,
the total Berry phase should be obtained from eq.\ (\ref{eq:single-berry})
by calculating the finite-temperature
canonical angular momentum density $S(r)$
composed of $u_j$, $v_j$, and $f(E_j)$
$\bigl($eq.\ (\ref{eq:ang-s})$\bigr)$.
   It is consistent with
an imaginary time path integral formulation.\cite{ao99}
   We should note here that the solutions of the BdG
equation
eq.\ (\ref{eq:bdg}),
i.e., the Bogoliubov wave functions $(u_j,v_j)$,
have {\it all the information on the system} on both the condensate
and non-condensate in a two-fluid picture.

   In Fig.\ \ref{fig:1}, we show $S(r)$
at several temperatures.
   $S(0)$ remains zero at finite temperatures
in Fig.\ \ref{fig:1}
due to the form of eq.\ (\ref{eq:ang-s})
and the necessary structure of $u_{n,l}(r)$ and $v_{n,l}(r)$
expanded by the Bessel functions.
   Because $S(0)=0$, even at finite temperatures
the transverse force does not depend on the vortex core
in the single vortex case.
   The particle density,~\cite{haya-j}
$\rho(r)=2 \sum_{E_j>0}
\bigl[|u_j (r)|^2 f(E_j) + |v_j (r)|^2 \{1-f(E_j)\}\bigr]$,
is simultaneously presented in Fig.\ \ref{fig:1}.
   Here, we adjust the chemical potential $\mu$ at each temperature
so that the particle density $\rho$ far from the vortex core is
invariable.~\cite{haya-j}
   With increasing temperature, $|S|$ decreases
with respect to $\rho /2$.
   We can find from Fig.\ \ref{fig:1} that the decrease of $|S|$
far from the core
obeys the temperature dependence of the Yosida function.~\cite{parks}
This result gives numerical proof of the physical estimation
in ref.~\onlinecite{ao99}.
   Then,
$|S|$ indicates the superfluid density $\rho_s /2$ at finite temperatures.

   Let us consider the two-vortex problem,
which is the main aim in the present study.
   We consider two vortices, labeled 1 and 2, moving with finite
relative velocity.
   Their relative dynamics is of interest here.
Hence, without loss of generality,
vortex 1 is at rest at position ${\bf r}_1$ and
vortex 2 at ${\bf r}_2(t)$ moves around vortex 1.
   We assume the vortices are separated so that
there is almost no overlap between the vortex bound states of each vortex.
   Then, wave functions bounded inside each vortex core are defined
separately from each other.
   A two-vortex pair potential is expressed in the product form,
$\Delta({\bf r},t)=
\Delta_v({\bf r}-{\bf r}_1)\Delta_v({\bf r}-{\bf r}_2(t))$,
where $\Delta_v({\bf r})=\Delta(r)\exp[-i\theta]$ denotes the pair potential
of the single vortex.
   The wave functions near each vortex are expected to be
close to those of the single vortex which have cylindrical symmetry,
but a phase factor is attached so that the above product form of
the pair potential is constructed from those wave functions.

\begin{figure}
\epsfxsize=77mm
\hspace{5.5mm}
\epsfbox{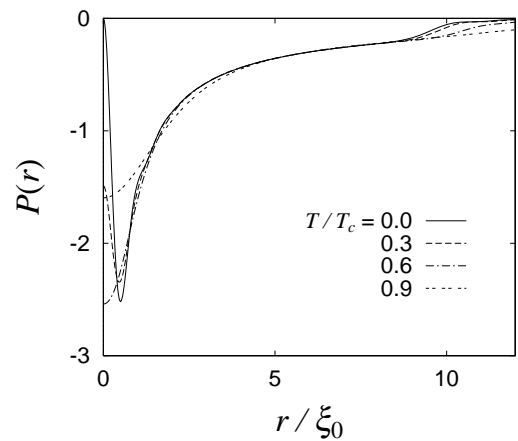}
\vspace{5.5mm}
\caption{
        Plot of $P(r)$ (in arbitrary units) obtained using exact
     wave functions of the vortex bound states around a vortex (see text).
        $k_{\rm F}\xi_0=4$.
}
\label{fig:2}
\end{figure}

   Under the above conditions, we obtain the bound-state contribution
to the Berry phase~\cite{dziarmaga,haya-u} as
$\Gamma_b=\Gamma_{1b}+\Gamma_{2b}$; the contribution from
the bound-state wave functions of vortex 1 is
$\Gamma_{1b}=-2\pi A \int dt
(R_x {\dot R}_y - R_y {\dot R}_x) / |{\bf R}|^2$,
and that of vortex 2 is
$\Gamma_{2b}=-2\pi A \int dt
(R_x {\dot R}_y - R_y {\dot R}_x) / |{\bf R}|^2
- \pi S_b \int dt (x_2 {\dot y}_2 - y_2 {\dot x}_2)$,
where ${\bf R}={\bf r}_2(t)-{\bf r}_1$,
$A=\int r dr P(r)$, (the center of each vortex is $r=0$), and
\begin{eqnarray}
P(r) &=&
-2 \sum_{0<E_{n,l}<\Delta(T)} \frac{1}{2}\Bigl[
|u_{n,l}(r)|^2 f(E_{n,l}) \nonumber \\
     & & {}+|v_{n,l}(r)|^2 \{1-f(E_{n,l})\}
\Bigr].
\label{eq:P}
\end{eqnarray}
   The second term in $\Gamma_{2b}$ is
the bound-state contribution to
the Berry phase
which leads to the Magnus force
in the absence of vortex 1,
and is zero ($S_b=0$) as seen in the single vortex case.
   What we aim to show in this letter is
that the bound-state contribution $P(r)$ is actually nonzero.
   In the previous two-vortex discussion,~\cite{dziarmaga}
this ``$|u_j|^2-|v_j|^2$ type"
contribution of the bound states has been disregarded
because of the equality $|u_j|=|v_j|$ for approximated bound-state
solutions.
   However, exact solutions in general must have essential asymmetry
between $|u_j|$ and $|v_j|$
due to the existence of the vortex itself,
which leads to a local breaking of particle-hole symmetry in
the density of states inside the vortex core~\cite{haya}
and an electric charging of the core.~\cite{haya-j}
   The Fermi functions should also be attached to
the Bogoliubov wave functions
on the analogy of $S(r)$
in eq.\ (\ref{eq:ang-s}).
   In Fig.\ \ref{fig:2}, we show
$P(r)$ which is calculated
with the exact solutions $(u_j,v_j)$ around the vortex,
and find it is certainly nonzero.

   On the other hand,
the wave functions of the extended states above the gap $\Delta(T)$
spread over vortices 1 and 2.
   To consider their contribution,
let us apply a discussion by Dziarmaga~\cite{dziarmaga}
based on an analogy of the impurity effect on the Magnus force
due to Ao and Thouless.~\cite{ao}
   The Berry phase is proportional to the mean number of superfluid particles
enclosed by a closed path around which a vortex is moved
adiabatically.~\cite{haldane}
   The stationary vortex 1, as a kind of impurity, traps particles
and excludes them from the extended states relevant to the Berry phases
picked up by the extended wave functions.
   A quantity which becomes a central issue is
$\delta S_{e}(r) = S_{e}(r) +\rho_s/2$,~\cite{dziarmaga}
the deviation of $S_e$ from its asymptotic value $-\rho_s/2$,
where $S_e(r)$ is the contribution from the extended states to
$S(r)$ in eq.\ (\ref{eq:ang-s}) around the stationary vortex 1.
   Then $2\int dr 2\pi r \delta S_e(r)$ ($\equiv B$) corresponds to
the excluded particle number
which works to decrease
the magnitude of
the single-vortex Magnus force
$|-h\rho_s/2|$ in spatial average.~\cite{dziarmaga}
   The contribution to the Berry phase is
$\delta \Gamma=-B \int dt (R_x {\dot R}_y - R_y {\dot R}_x) / |{\bf R}|^2$.

   Taking the variation of the Berry phase
$\Gamma_b +\delta \Gamma$
about ${\bf r}_2$,
we finally obtain the force ${\bf F}_2$
acting on vortex 2
due to the presence of vortex 1,
${\bf F}_2 = \bigl( h (4 A + B/\pi)/|{\bf R}|^2 \bigr)
\bigl[{\hat z}\times{\dot {\bf R}}
+ {\hat z}\cdot({\bf R}\times{\dot {\bf R}}) {\bf R}/|{\bf R}|^2 \bigr]$,
and we propose to add it to the general form~\cite{ao95} of
the usual vortex equation of motion.
   In addition to the nondissipative transverse force
(the first term in large brackets),
we obtain an interesting dissipative
term (the second one) which, to our knowledge, has not been reported so far.
   An analysis of this dissipative term is
a future task.
   In what follows, we discuss the nondissipative transverse force.

\begin{figure}
\epsfxsize=77mm
\hspace{5.5mm}
\epsfbox{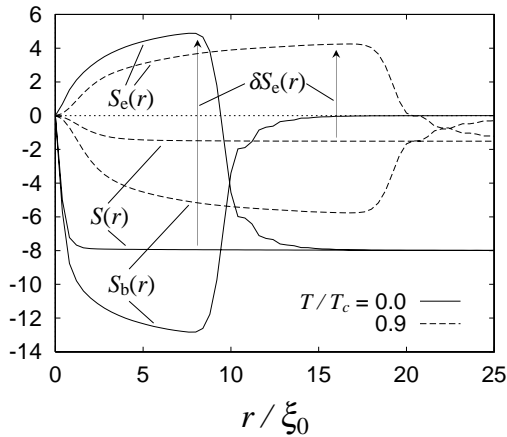}
\vspace{5.5mm}
\caption{
        To the canonical angular momentum density $S(r)$
     around a vortex (in arbitrary units),
     the contribution from the vortex bound states
     $S_b(r)$ and
     that from the extended states
     $S_e(r)$,
     where $S(r)=S_b(r)+S_e(r)$.
     The deviation $\delta S_e(r)=S_e(r)+\rho_s /2$ (see text).
        $k_{\rm F}\xi_0=4$.
}
\label{fig:3}
\end{figure}

   In Fig.\ \ref{fig:3}, we show our numerical results for
$S_e(r)$ together with the bound-state contribution $S_b(r)$
and total $S(r)$.
   The deviation $\delta S_{e}(r)$ is positive
as noted in ref.~\onlinecite{dziarmaga}.
   It is of importance to note here that
the deviation can become
$\delta S_{e}(r)>|-\rho_s/2| \bigl(=|S(\infty)| \bigr)$ locally
around the vortex core, as seen in Fig.\ \ref{fig:3}.
   We are able to understand this by the fact~\cite{gygi} that
in general
the extended states contribute
in terms of negative rotation to the supercurrent
around a vortex,
and the bound states contribute in terms of positive rotation to it.
   We note in Fig.\ \ref{fig:3} that
at a higher temperature, the region of such large $\delta S_{e}(r)$
expands following the divergence of
the coherence length near $T_{\rm c}$.
   Accordingly, near $T_{\rm c}$
the region around such stationary vortices
where $\delta S_{e}(r)>|-\rho_s/2|$,
can predominantly occupy the inside of, e.g., a superconductor,
and thus the transverse force acting on the moving vortex
can become not only a decreased Magnus force but also an opposite force.
   It may naturally explain the sign-reversal Hall effect
near $T_{\rm c}$ observed experimentally.~\cite{hagen,nagaoka}
   The detailed structure of $\delta S_{e}(r)$
can depend on the details of various vortex cores, and thus
on the doping~\cite{himeda} in high-$T_{\rm c}$ cuprates.

   Effects of surrounding vortices
on a vortex
are usually taken into account
through the superfluid velocity dependent part of the Magnus force,
$\beta{\hat z} \times {\bf v}_s$, where
${\bf v}_s$ is the superfluid velocity including
the contribution of the surrounding vortices.
   In the present study, in contrast, we proposed quite a different force
due to the surrounding vortices, $\alpha{\hat z} \times {\bf v}_L$,
which acts on a moving vortex.
   According to Newton's action-reaction principle,
our transverse forces acting on the moving vortex
simultaneously act on the surrounding vortices in each opposite
direction.
   If the surrounding vortices are pinned ones,
our force should be included in a depinning condition for those vortices.

   In conclusion, we considered how the details of the vortex core
can affect the vortex dynamics by the many-body vortex effect
other than the possibility of the spectral flow force
which is due to $\tau$.
   The present study throws a light on the possibility for
the transverse force which may depend on
the details of vortex cores,
without abandoning the robust Magnus force theory
for the single vortex.

   A numerical simulation adopting the robust Magnus force and the force
${\bf F}_2$, and a direct two-vortex system analysis by
numerically calculating the BdG for a system containing two vortices,
would be interesting and are left for future work.

I am grateful to Ping Ao for bringing my attention to his work
and for helpful discussions.


\end{document}